\documentclass[3p,times,singlecolumn]{elsarticle}

\usepackage{graphicx,stfloats} 
\usepackage{color}
\usepackage{gensymb}
\usepackage{subcaption}
\usepackage{float}
\usepackage{comment}
\usepackage{booktabs}
\usepackage[colorlinks=true, allcolors=blue]{hyperref}
\usepackage[export]{adjustbox}%
\usepackage{amsmath}
\usepackage{amssymb}
\usepackage{amsthm}
\usepackage{bbding}
\usepackage{pifont}
\usepackage{wasysym}
\usepackage{amsmath}
\biboptions{sort&compress}

\journal{}

\usepackage[dvipsnames]{xcolor}
\usepackage[normalem]{ulem}
\usepackage{tikz}
\usepackage{pgfplots}
\pgfplotsset{compat=1.17}
\usepackage{cancel}

\usepackage{bm}
\newcommand{\q}{\bm{q}}
\newcommand{\Q}{\bm{Q}}
\newcommand{\Ss}{\bm{S}}
\newcommand{\x}{\bm{x}}
\newcommand{\w}{\bm{w}}
\newcommand{\W}{\bm{W}}
\newcommand{\vs}{\vspace{0mm}}
\newcommand{\vsv}{\vspace{0mm}}
\newcommand{\Psib}{\bm{\Psi}}
\newcommand{\Lambdab}{\bm{\Lambda}}
\newcommand{\Thetab}{\bm{\Theta}}

\begin{document}
\begin{abstract}
We demonstrate that accurate computation of the spectral proper orthogonal decomposition (SPOD) critically depends on the choice of frequency resolution. Using both artificially generated data and large-eddy simulation data of a turbulent subsonic jet, we show that the optimal choice depends on how rapidly the SPOD modes change in space at adjacent frequencies. Previously employed values are found to be too high, resulting in unnecessarily biased results at physically important frequencies. A physics-informed adaptive frequency-resolution SPOD algorithm is developed that provides substantially less biased SPOD modes than the standard constant resolution method. 
\end{abstract}

\begin{frontmatter}

\title{Optimal frequency resolution for spectral proper orthogonal decomposition}

\author[label1]{Liam Heidt\footnote{\noindent Corresponding author: lheidt@caltech.edu}}
\author[label2]{Tim Colonius}

\address[label1]{Graduate Aerospace Laboratories of the California Institute of Technology, California Institute of Technology, California, USA}
\address[label2]{Department of Mechanical and Civil Engineering, California Institute of Technology, California, USA}

\end{frontmatter}
\ifdefined \wordcount
\clearpage
\fi

\section{Introduction} \vsv
Spectral proper orthogonal decomposition (SPOD) \citep{lumley1967,lumley2007stochastic} is a data-driven method used to extract coherent structures from statistically stationary data and has been extensively used to study flow physics \citep{schmidt2018spectral,araya2017transition,abreu2020spectral} and to develop reduced order models \citep{chu2021stochastic,cammilleri2013pod}.
SPOD is typically estimated using the Welch periodogram method, which requires the choice of frequency bin width consistent with the amount of available data. Modes are subsequently computed using frequency-domain data averaged over this bin width. This frequency resolution is crucial due to the well-known bias-variance tradeoff \citep{schmidt2020guide}, and this resolution is typically chosen such that spectral peaks, if any, are sufficiently resolved. 

This work demonstrates that the choice of frequency resolution also depends on how rapidly the SPOD modes change {\it in space} from one frequency to the next. We first illustrate this using artificially generated data and then on a previously analyzed turbulent subsonic jet.  We show previously employed values of bin width are too high, resulting in unnecessarily biased results at physically important frequencies. We then develop a physics-informed adaptive frequency-resolution algorithm that provides substantially less biased SPOD modes than the standard constant resolution method. Lastly, we demonstrate that the alignment metric widely used to quantify the similarity between different SPOD modes (or between SPOD and resolvent analysis \citep{mckeon2010critical}) is sensitive to bias and the degree of statistical convergence, so care must be taken when inferring conclusions. \vs

\section{Computation of SPOD and the influence of $\Delta f$} \label{sec:SPODmethod} \vsv
An overview of SPOD is provided here; the reader is referred to \citep{lumley1967,towne2017spectral,schmidt2020guide} for details. 
Let $\q_k$ be a possibly complex vector-valued snapshot of a statistically stationary flow at time $t_k$ on the spatial domain $\Omega$. The snapshot length $N$ equals the number of variables multiplied by the number of spatial locations. Assume we have $N_t$ equally spaced snapshots available $t_{k+1} = t_k + \Delta t$. 
To employ Welch's method, we obtain the overlapping data matrix $\Q^{(n)} = [\q_1^{(n)}, \q_2^{(n)}, \cdots, \q_{N_s}^{(n)}]$, where $\q_k^{(n)} = \q_{k+(n-1)(N_s-N_0)}$, $N_s$ is the number of snapshots per block, $N_0$ is the number of snapshots that overlap, and $N_b$ is the number of blocks. Assuming ergodicity, each block is an independent flow realization. Next, the Fourier transform of each windowed block $\tilde{\Q}^{(n)} = {\Q}^{(n)} \cdot \w = [w_1\q_1^{(n)}, w_2\q_2^{(n)}, \cdots, w_{N_s}\q_{N_s}^{(n)}]$  is computed $\hat{\Q}^{(n)} = FFT(\tilde{\Q}^{(n)}, N_f)$, where $\w$ is a window that reduces spectral leakage and $FFT({\Q}^{(n)}, N_f)$ is the $N_f$ length Fourier transform of $\tilde{\Q}^{(n)}$. If $N_s < N_f$, the data is zero-padded. This computes SPOD at $N_f$ frequencies with a spectral resolution $\Delta f = 1/(N_s \Delta t)$, where the SPOD estimate at $f_k$ averages the data in $f = [f_k - \Delta f/2, f_k + \Delta f/2]$. The frequency vector is given by $f_k = \frac{k - 1}{N_f \Delta t},\ k = [1, N_f]$. Alternatively, a non-FFT method can be used, like the Goertzel \citep{goertzel1958algorithm} method, where specific discrete frequencies are chosen.  The cross-spectral density (CSD) at $f_k$ is estimated by $\Ss_{f_k} = \hat{\Q}_{f_k}\hat{\Q}_{f_k}^*$, where $\hat{\Q}_{f_k} = \sqrt{\kappa}\tilde{\Q}^{(n)}$, $\kappa = \Delta t/(||\w||_2 N_b)$, and $^*$ is the Hermitian transpose. SPOD is computed by
\begin{equation}
    \Ss_{f_k} \W \Psib_{f_k} = \Psib_{f_k} \Lambdab_{f_k}, 
\end{equation}
where $\W$ is a weighting matrix. The estimated SPOD modes are given by the columns of $\Psib_{f_k}$ and are ranked by their energy $\lambda_j$. The CSD matrix, $\Ss_{f_k}$, is an $N \times N$ matrix, far too large for typical problems. Thus, to practically compute SPOD, we employ the exact method-of-snapshot approach \citep{sirovich1987turbulence}, giving
\begin{subequations}
\begin{align}
    \hat{\Q}_{f_k}^* \W \hat{\Q}_{f_k} \Thetab_{f_k} = \Psib_{f_k} \Thetab_{f_k}, \\
    \Psib_{f_k} = \hat{\Q}_{f_k} \Thetab_{f_k} \Lambdab_{f_k}^{-1/2}.
\end{align}
\end{subequations}
This \textit{estimate} converges as the number of blocks $N_b$ and the number of snapshots in each block $N_s$ are increased together \citep{welch1967use,bendat2011random}. For a fixed amount of data $N_t$, as $N_s$ increases, $N_b$ decreases. This leads to an estimate with a greater variance but reduced bias. The vice-versa is also true. Practitioners attempt to determine a $N_s$ (and thus $N_b$) that results in the most accurate SPOD modes.  It is well known that for flows exhibiting a spectral peak, such as an open cavity, a greater $N_s$ is required to ensure minimal bias. This is because when computing SPOD at a discrete frequency $f_k$, the energy/data from $[f_k - \Delta f/2, f_k + \Delta f/2]$ is averaged. Thus, if the energy/flow in $[f_k - \Delta f/2, f_k + \Delta f/2]$ varies substantially, the discrete estimate of the SPOD modes (and energy) at $f_k$ will not accurately reflect the true modes.  \vs

\section{Artificial problem} \label{sec:dummyproblem} \vsv
In addition to the bias being high if $\Delta f$ is large compared to the bandwidth of a spectral peak, we hypothesize that a small $\Delta f$ is required when the SPOD modes change rapidly as a function of $f$. By that, we mean if the \textit{``true"} SPOD modes change substantially between $f_k \pm \Delta f/2$, then the estimated SPOD at $f_k$ will contain a high amount of bias (i.e. error). To verify this, we construct artificial data with a known SPOD spectrum/modes and then analyze the impact of $\Delta f$ as a function of how rapidly the known SPOD modes vary. 

By definition, a random statistically-stationary process $q(x, t)$ with a SPOD spectrum $\lambda_j(f)$ and modes $\psi_j(x, f)$ has a CSD, $S(x,x^\prime, f) = \sum_{j = 1}^\infty \lambda_j(f) \psi_j(x, f) \psi_j(x, f)^* $. Data with these properties can be generated by 
\begin{subequations}
\begin{align}
    q(x, t) &= \int_{-\infty}^\infty \hat{q}(x, f) e^{-i 2\pi f t} df \\
     &= \int_{-\infty}^\infty \tilde{F}\Big\{G\Big(0, S(x, x^\prime, f)\Big), f\Big\}\ e^{-i 2\pi f t} df, 
\end{align}
\end{subequations}
where $g(x, t) = G(0, S(x, x^\prime, f))$ is zero-mean Gaussian white noise with a covariance kernel of $S(x, x^\prime, f)$ (i.e. a covariance kernel equal to the CSD of $q(x, t)$ at frequency $f$) and $\tilde{F}\big\{g(x, t), f^\prime\big\}$ is a Kronecker delta filter such that $\hat{g}(x, f) = 0\ \forall f \ne f^\prime$. 
The dominant SPOD modes are defined as 
\begin{equation}
    \psi_1(x, f) = e^{-x^2/(2 c_1^2)} e^{1i x fc_2}/(\pi^{1/4} c_1^{1/2}),
\end{equation}
where $\pi^{1/4} c_1^{1/2}$ is a normalization constant ensuring $\langle \psi_1(x, f), \psi_1(x, f) \rangle = 1$, $c_1$ is a constant that defines the spatial distribution, and $c_2$ defines how rapidly the modes change from one frequency to the next. 
Increasing $c_2$ results in more rapidly changing SPOD modes. Additionally, $\lambda_1(f) = 1, \lambda_{2}(f) = 0.5$, and $\psi_{2}(x, f) = r(x, f)$, where $r(x, f)$ is a randomly spatially distributed mode orthogonal to $\psi_1(x, f)$. 

Next, we generate discrete data using our analytical SPOD modes, numerically compute SPOD using varying $\Delta f$, and then compute the alignment between the numerical (n) and analytical (a) SPOD modes $\alpha = \langle \psi_{1, \text{a}}(x, f), \psi_{1, \text{n}}(x, f) \rangle$. Since this process is stochastic, we repeat this process to obtain a converged averaged alignment. In this artificial case, each frequency is independent, allowing us to also average over different frequencies. For all results, we confirm sufficient convergence (not shown). We employ an equally spaced grid $x = [-20, 20]$ with $N_x = 801$, $\Delta t = 1$, $c_1 = 5$, $c_2 = 10, 20, 40, 80, 160$, $N_t = 5000, 10000, 20000, 40000, 80000, 160000$, and an SPOD block overlap of $67\%$. 

\begin{figure}[htb!]
\centering
\begin{subfigure}[b]{0.41\textwidth}
\includegraphics[width=\textwidth]{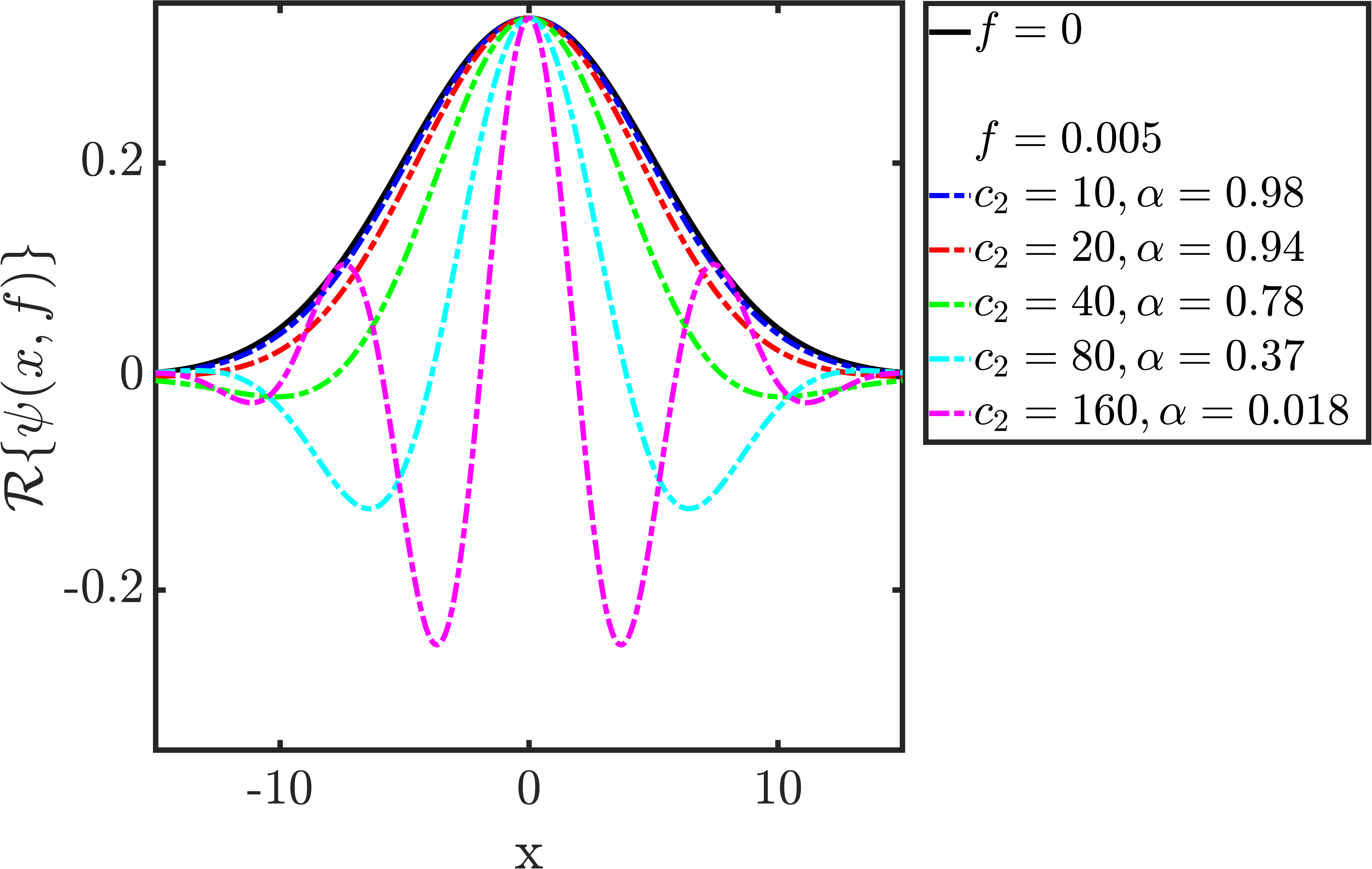}\llap{
  \parbox[b]{2.5in}{a)\\\rule{0ex}{1.5in}
  }}
\end{subfigure}  \hspace{0.75mm}
\begin{subfigure}[b]{0.28\textwidth}
\includegraphics[width=\textwidth]{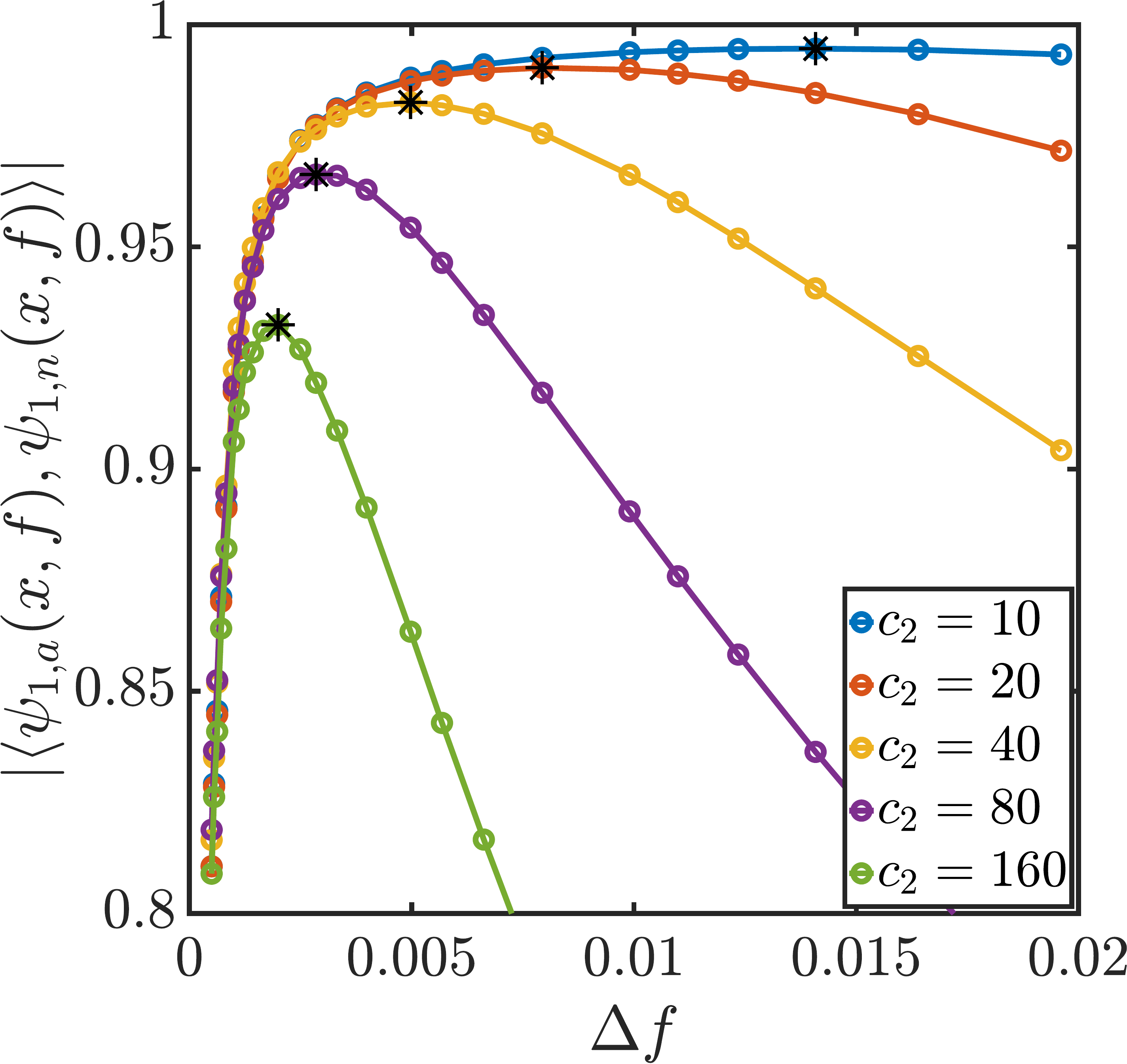}\llap{
  \parbox[b]{1.7in}{b)\\\rule{0ex}{1.5in}
  }} 
\end{subfigure} \hspace{0.75mm}
\begin{subfigure}[b]{0.28\textwidth}
\includegraphics[width=\textwidth]{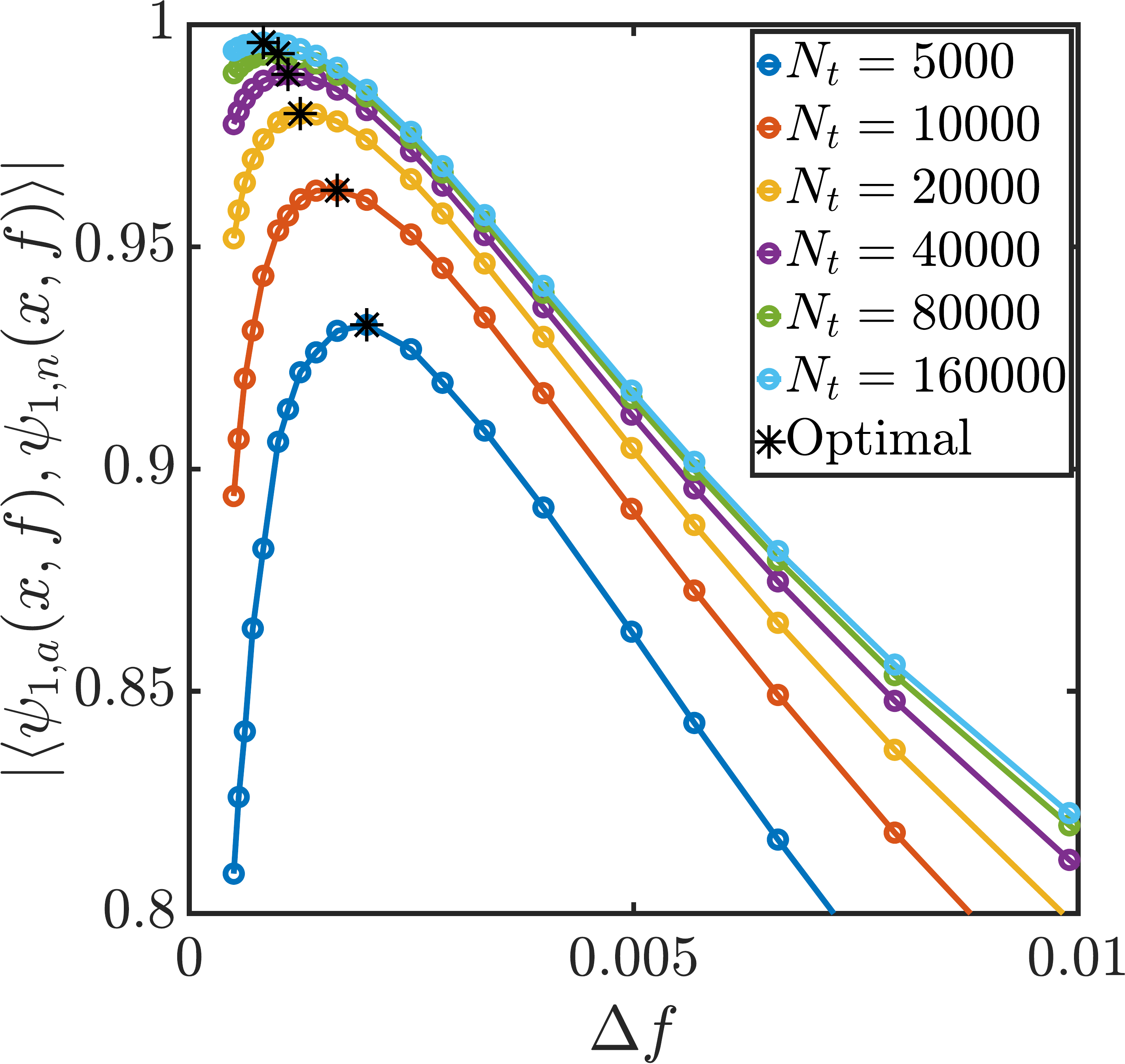}\llap{
  \parbox[b]{1.7in}{c)\\\rule{0ex}{1.5in}
  }} 
\end{subfigure}
\caption{Real component of the analytical artificial problem SPOD modes (a) and average alignment between numerical and analytical SPOD modes as a function of $\Delta f$ for varying $c_2$ (b) and $N_t$ (c). }
\label{fig:dummyoverall}
\end{figure}

In figure \ref{fig:dummyoverall} (a), we display an example mode $\mathcal{R}\{\psi_1(x, f)\}$ at $f = 0, 0.01$ and the alignment $\alpha = |\langle \psi_{1}(x, 0), \psi_{1}(x, 0.01) \rangle|$. We see that as $c_2$ increases, the modes vary more rapidly. In figure \ref{fig:dummyoverall} (b), we display the average alignment between the analytical and numerical SPOD modes as a function of $\Delta f$ for the different values of $c_2$ and $N_t = 5000$. For a given $c_2$, we find that as $\Delta f$ increases, the alignment first increases and then decreases. This is the standard bias-variance tradeoff with high variance at low $\Delta f$ and high bias at high $\Delta f$. We clearly note that as $c_2$ increases, the optimal $\Delta f$ decreases, confirming our hypothesis that rapidly changing SPOD modes require a smaller $\Delta f$. In figure  \ref{fig:dummyoverall} (c), we show the average alignment for a constant $c_2 = 160$ as a function of $N_t$. As expected, the error decreases for larger $N_t$, and the optimal $\Delta f$ also decreases for increasing $N_t$. Here, we find that if a too-high $\Delta f$ is chosen, the error does \textit{not} decrease with increasing $N_t$ (since the error is dominated by bias and not variance). Thus, selecting an appropriate $\Delta f$ is crucial, and an appropriate value varies greatly depending on how rapidly the SPOD modes vary. 

\section{Turbulent jet} \vsv
We now perform a similar investigation on an LES of an isothermal Mach 0.4 round jet with a Reynolds number $Re_j = \rho_jU_jD/\mu_j = 4.5\times 10^5$, where $\rho_j, U_j, D, \mu_j$ are the jet density, velocity, diameter, and dynamic viscosity. For details, we refer the reader to \citep{bres2018importance,schmidt2018spectral}. We simulate $t_{sim}c_{\infty}/D = 16000$ flow through times saving $N_{tot} = 80000$ snapshots ($4 - 8\times$ previous studies of similar turbulent jets \citep{schmidt2018spectral,heidt2021analysis,nekkanti2022triadic}), with $\Delta t = 0.2$. We look at the axisymmetric fluctuating components only. 

We now compute SPOD using a variety of $N_s$ and $N_t$ for a range of $f$ (using either method described in \S \ref{sec:SPODmethod}). Since the true SPOD modes are not known, we compare them to resolvent analysis modes generated using an identical method as in \citet{heidt2021analysis}. While resolvent analysis modes are not exactly SPOD modes for turbulent jets, the alignment has been shown to be high over the range of important frequencies. In addition, any error due to bias or variance is not likely to align in the direction of the resolvent mode. Thus, improvement in the alignment can be considered improvements to the accuracy of the corresponding SPOD mode. 

\begin{figure}[htb!]
\centering{
\begin{subfigure}[b]{0.35\textwidth}
\includegraphics[height=1\textwidth]{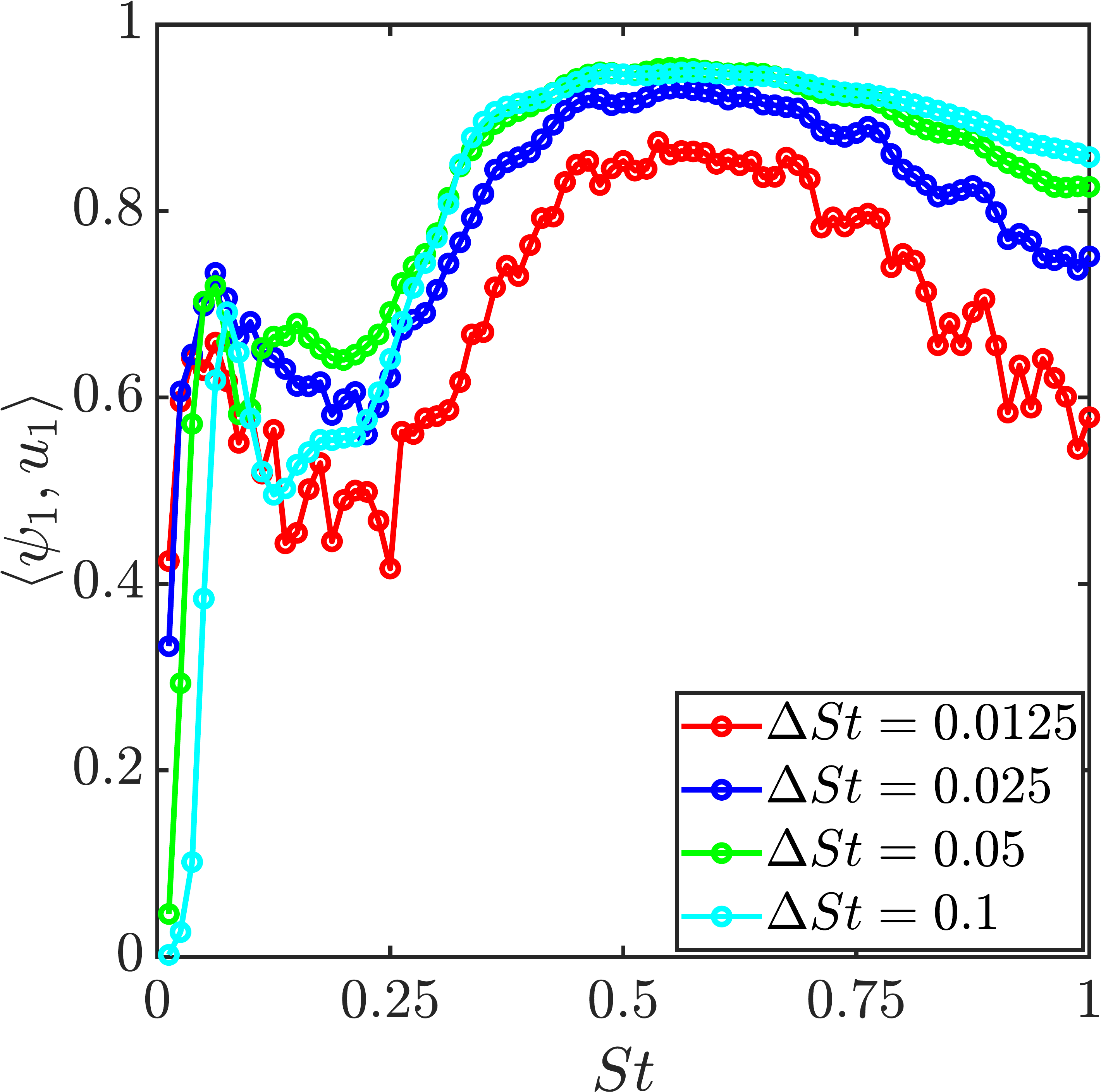}\llap{
  \parbox[b]{2.3in}{a)\\\rule{0ex}{2.05in}
  }}   
\end{subfigure}  \hspace{10mm}
\begin{subfigure}[b]{0.35\textwidth}
\includegraphics[height=1\textwidth]{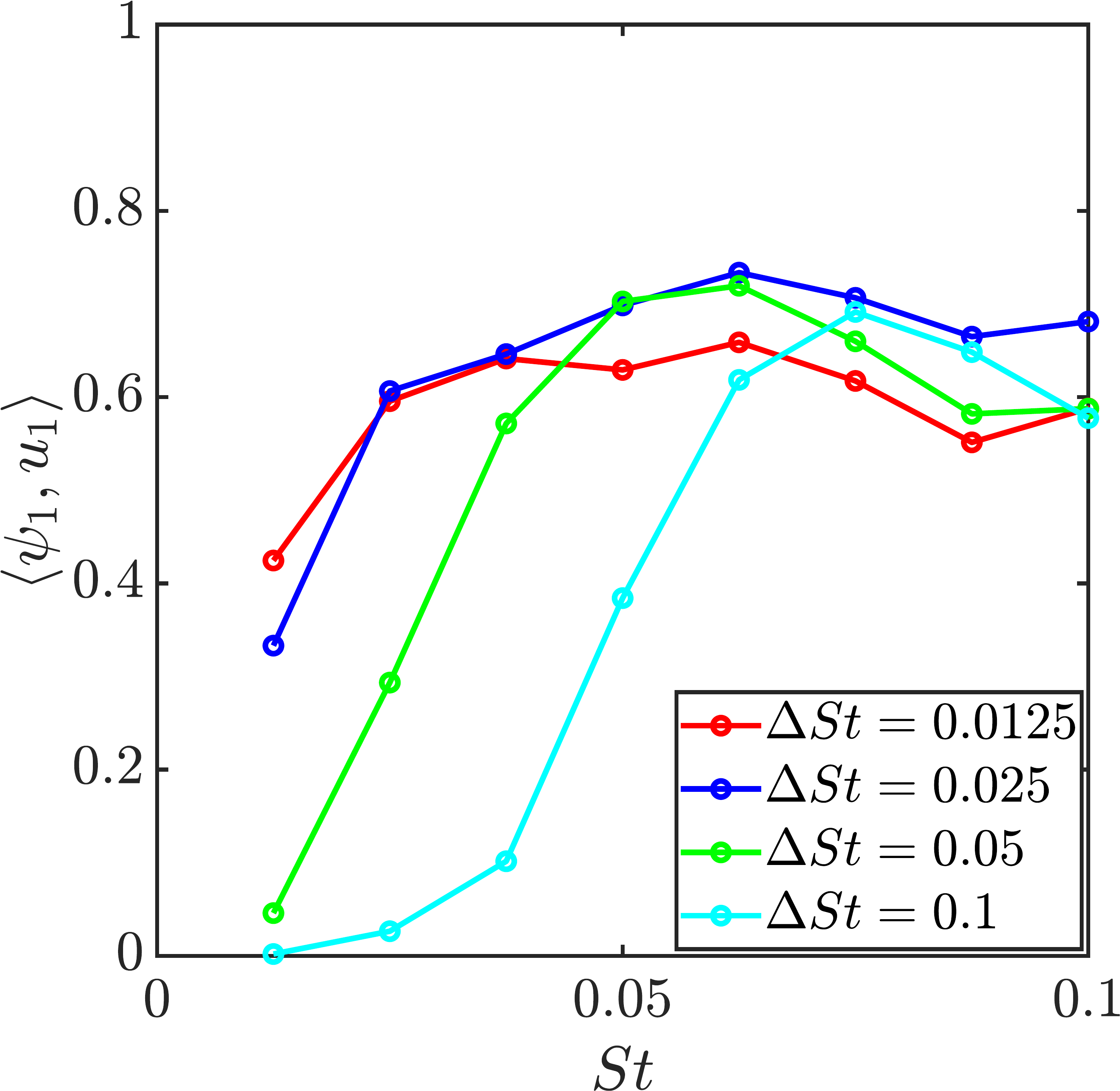}\llap{
  \parbox[b]{2.3in}{b)\\\rule{0ex}{2.05in}
  }}     
\end{subfigure}}
\caption{Alignment between resolvent and SPOD for $N_t = 10000$ for various $\Delta St$ (a) (zoomed in (b)).}
\label{fig:jetoverall1}
\end{figure}

\begin{figure}[htb!]
\centering
\includegraphics[width=0.373333333\textwidth]{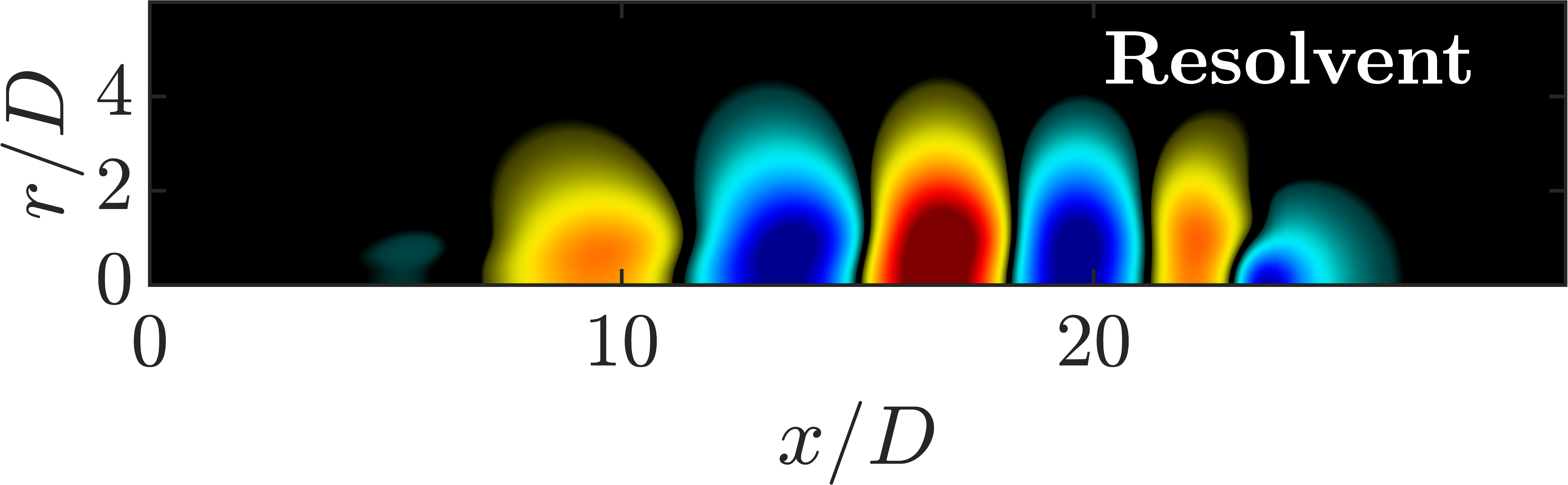} \\
\hspace{0.0mm} \includegraphics[width=0.373333333\textwidth,valign=t]{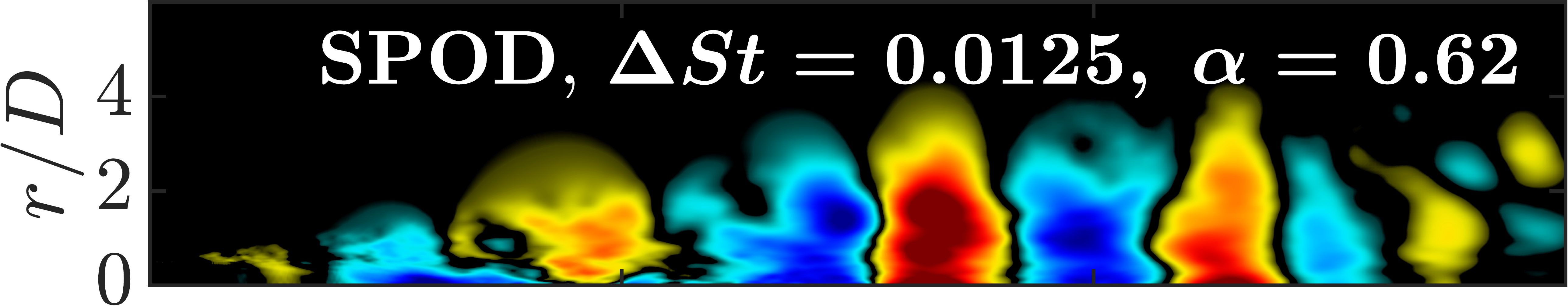}
\hspace{0.8mm}\includegraphics[width=0.338207593\textwidth,valign=t]{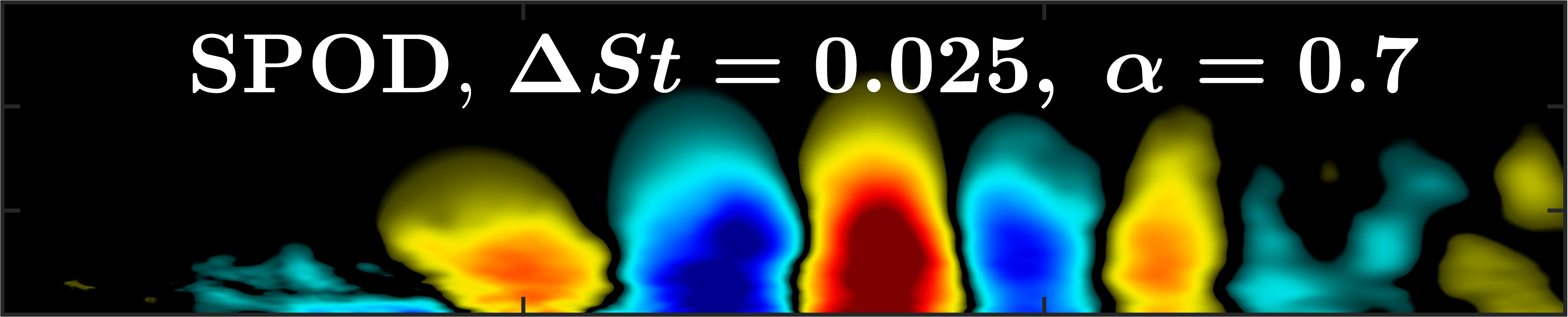}\\
\hspace{0.5mm} \includegraphics[width=0.373333333\textwidth,valign=b]{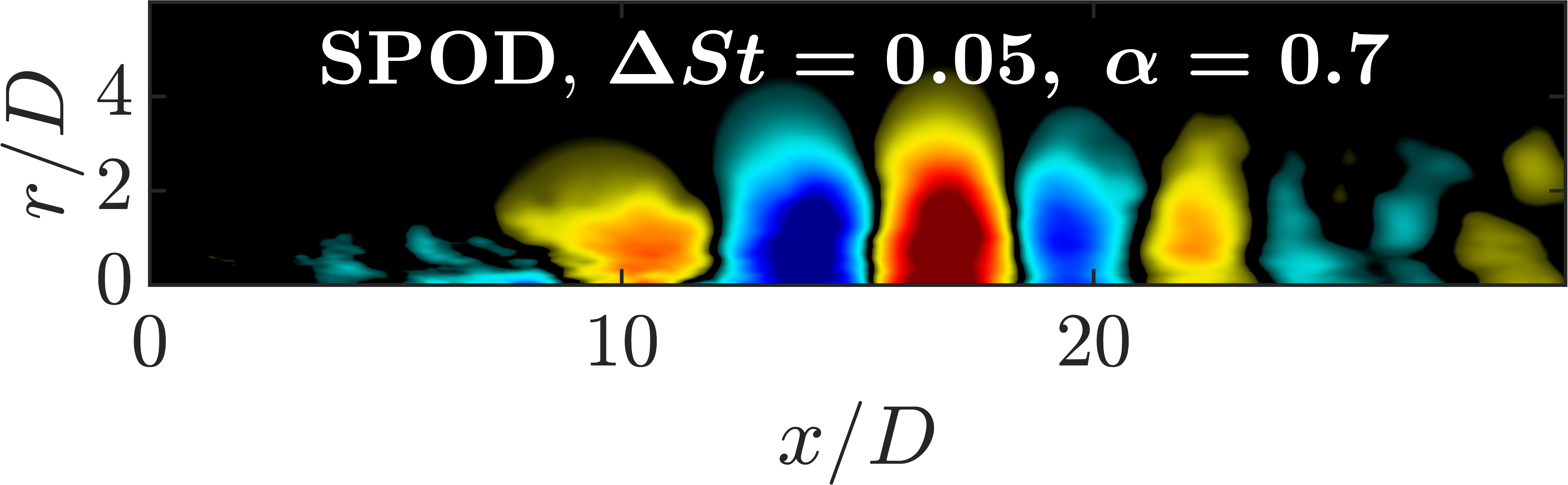} 
  \includegraphics[width=0.341687542\textwidth,valign=b]{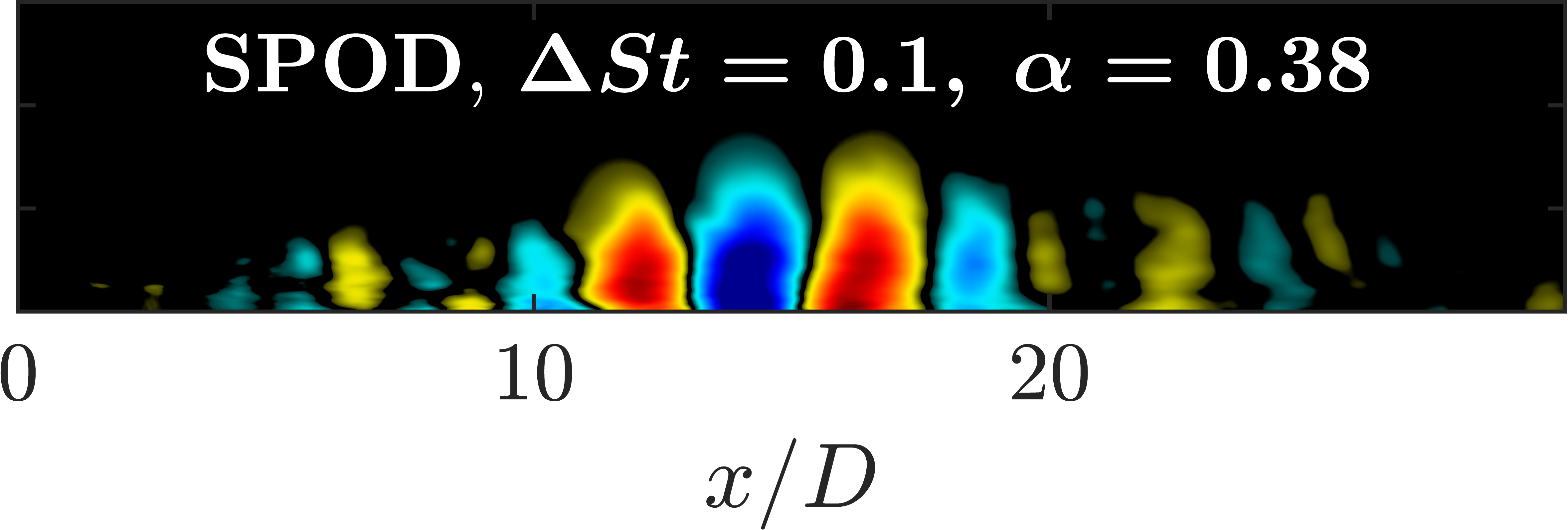} 
\caption{Dominant resolvent and SPOD modes at $St = 0.05$ for several $\Delta St$ at $St = 0.05$.}
\label{fig:jetoverall1modes}
\end{figure}

In figure \ref{fig:jetoverall1} (a, b), we display the alignment between the $N_t = 10000$ SPOD case for several $\Delta St$. For $N_t < N_{tot}$, multiple sets of SPOD are computed with $50\%$ overlap between datasets, which are used to compute the average alignment.  We see that for higher $St$ $(> 0.5)$, increasing $\Delta St$ results in better alignment (i.e. decreases variance). For low $St$ $(\approx0)$, increasing $\Delta St$ decreases alignment due to increasing bias. Using a constant $\Delta St = 0.05$ (a typical value when studying subsonic jets) results in severely biased results when investigating $St \rightarrow 0$, which is an important regime that contains more energy than any other frequencies. In the intermediate region $(St \in [0.1, 0.5])$, the alignment initially increases with increasing $\Delta St$ due to decreasing variance and then decreases due to increasing bias. This occurs because the wavelength of the dominant mode scales approximately inversely with $St$ due to the constant phase speed of the wavepackets. Thus, modes at $St = 0.05, 0.075$ (resolvent mode alignments $|\langle u_1(St=0.05), u_1(St=0.075) \rangle| = 0.4$) vary more rapidly than modes at $St = 1.5, 1.525$ ($|\langle u_1(St=1.5), u_1(St=1.525) \rangle| = 0.85$). Thus, based on \S \ref{sec:dummyproblem}, we expect a lower/higher $\Delta St$ to be required at low/high $St$, respectively. In figure \ref{fig:jetoverall1modes}, we display the dominant resolvent and SPOD modes at $St = 0.05$ for $\Delta St = 0.0125, 0.025, 0.05$. Here, we find that as $\Delta St$ increases, the mode initially becomes less noisy, indicating less variance. In contrast, the mode shape changes substantially for $\Delta St = 0.1$, with the wavelength greatly reducing due to increasing bias.

\begin{figure}[ht]
\centering
\begin{subfigure}[b]{0.358\textwidth}
\includegraphics[width=\textwidth]{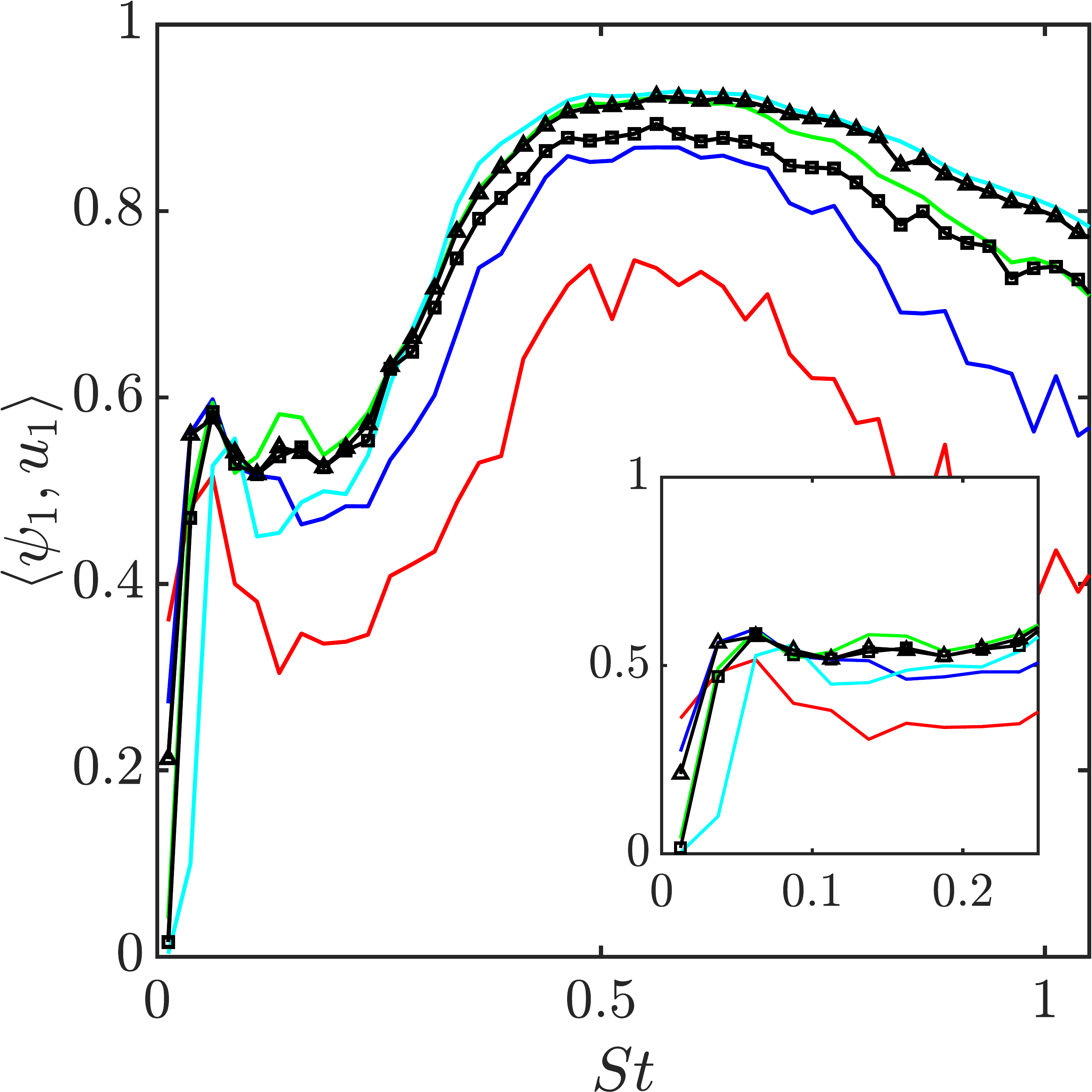}\llap{
  \parbox[b]{1.9in}{a)\\\rule{0ex}{2.05in}
  }}       
\end{subfigure} 
\begin{subfigure}[b]{0.3108\textwidth}
\includegraphics[width=\textwidth]{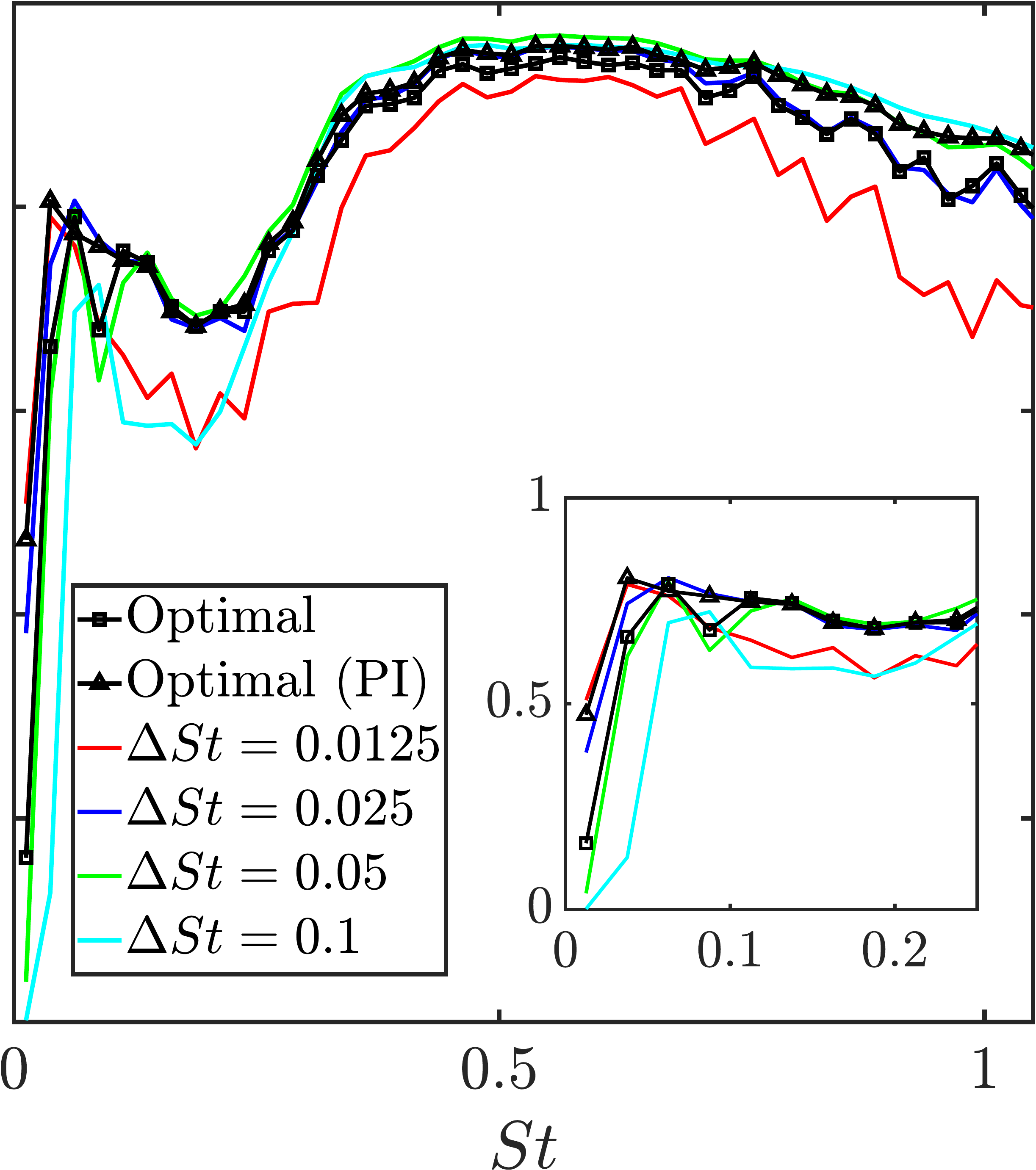}\llap{
  \parbox[b]{1.9in}{b)\\\rule{0ex}{2.05in}
  }}         
\end{subfigure} 
\begin{subfigure}[b]{0.3138\textwidth}
\includegraphics[width=\textwidth]{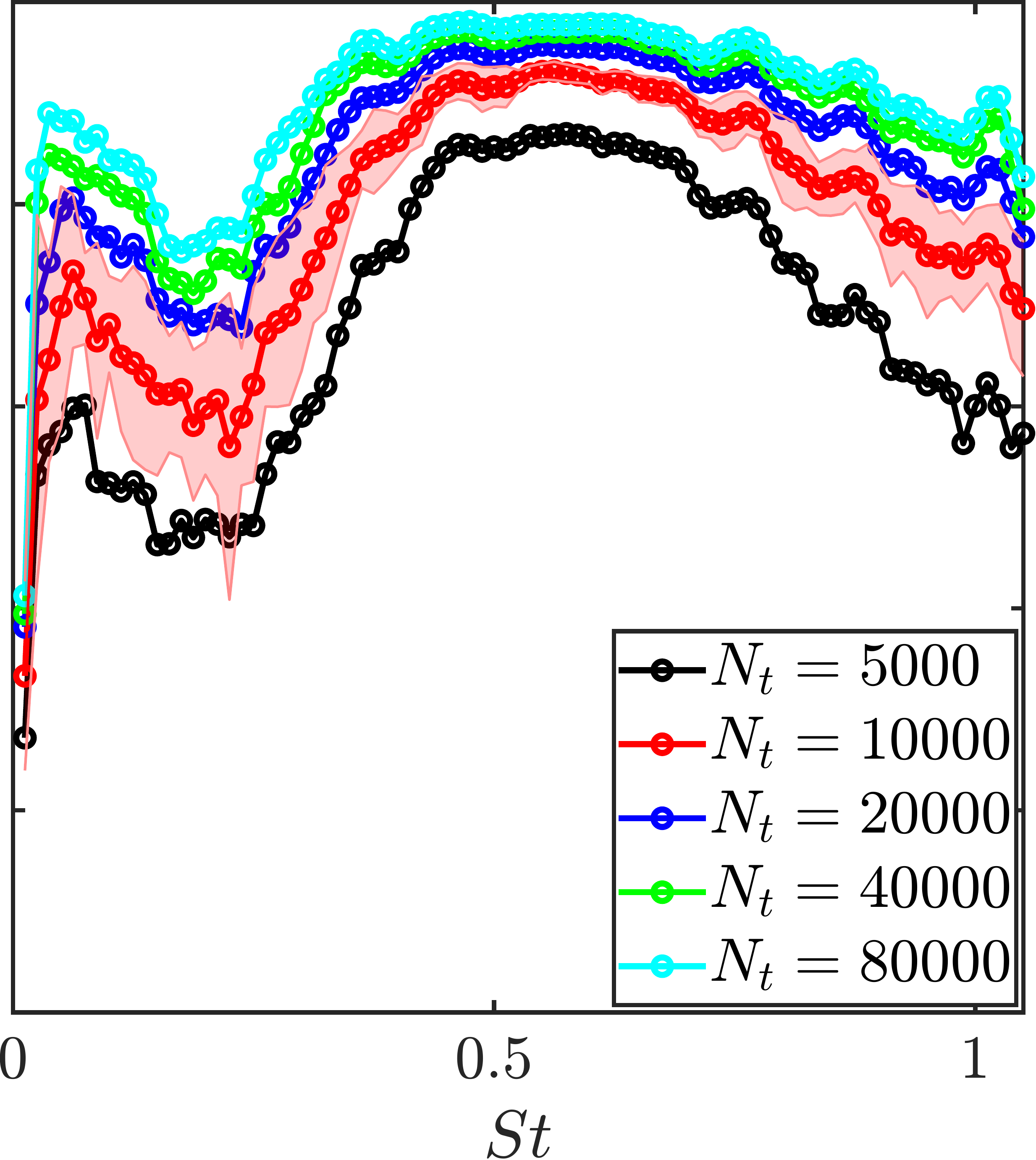}\llap{
  \parbox[b]{1.9in}{c)\\\rule{0ex}{2.05in}
  }}         
\end{subfigure} \hfill
\caption{Alignment between dominant SPOD and resolvent mode for $N_t = 5000$ (a), $N_t = 20000$ (b) jet for varying $\Delta St$ and the two adaptive methods. (c) Alignment between dominant SPOD and resolvent mode for $\Delta St = 0.025$ for varying $N_t$ along with standard deviation of alignment for $N_t = 10000$ (shaded region).}
\label{fig:jetoptimal}
\end{figure}

We now create an adaptive resolution SPOD algorithm to vary $\Delta f$ for improved alignment. We employ a similar cost function as \citet{yeung2023adaptive} and begin with a small $\Delta f_j$ and increase $\Delta f_{j+1}$ until 
\begin{equation}
    \mathcal{J} = 1 - |\langle \psi_{1,\Delta f_j}(f_k),  \psi_{1,\Delta f_{j+1}}(f_k) \rangle| < e, 
\end{equation}
where $\psi_{1,\Delta f}(f_k)$ is the dominant SPOD mode at frequency $f_k$ computed using a frequency bin width of $\Delta f$ and $e$ is a convergence tolerance. We also find that at frequencies where the SPOD modes vary slowly, a better convergence can be reached for a fixed amount of data (as seen in \S \ref{sec:dummyproblem}). Thus, we vary $e = e(f)$ using known physics. For jets, due to the constant phase speed, increasing frequency results in more slowly varying SPOD modes. Thus, we assume $e(f) = e_2/f$, such that 
\begin{equation}
    \mathcal{J} = 1 - |\langle \psi_{1,\Delta f_j}(f_k),  \psi_{1,\Delta f_{j+1}}(f_k) \rangle| < e_2/f_k,
\end{equation}
In figure \ref{fig:jetoptimal} (a, b), we show the alignment between the dominant SPOD and resolvent modes for several $\Delta St$ and the two adaptive procedures described, using a convergence constant of $e = 0.5, e_2 = 0.1$, for $N_t = 5000, 20000$. Here, both adaptive methods switch from a lower $\Delta St$ at low $St$ to a higher $\Delta St$ at higher $St$. Using a fixed convergence tolerance results in the optimal $\Delta St$ decreasing with increasing data lengths (consistent with results found in $\S$ \ref{sec:dummyproblem} since the variance decreases and $\Delta St$ becomes smaller to reduce the bias). Both adaptive methods result in greatly improved SPOD modes compared to employing a fixed $\Delta St$. In particular, the modified adaptive method provides excellent alignment across all frequencies and data lengths. 

In general, we note that no value of $e, e_2$ can be considered universal, and for each case, practitioners must use physical intuition to determine the best values of $\Delta f$. In general, we recommend increasing $\Delta f$ while the mode shape stays similar but becomes less \textit{noisy} (i.e. decreasing variance). If the mode shape changes, $\Delta f$ is likely too high, and large bias errors will occur. 

\begin{figure}[htb!]
\centering
\includegraphics[width=0.4\textwidth]{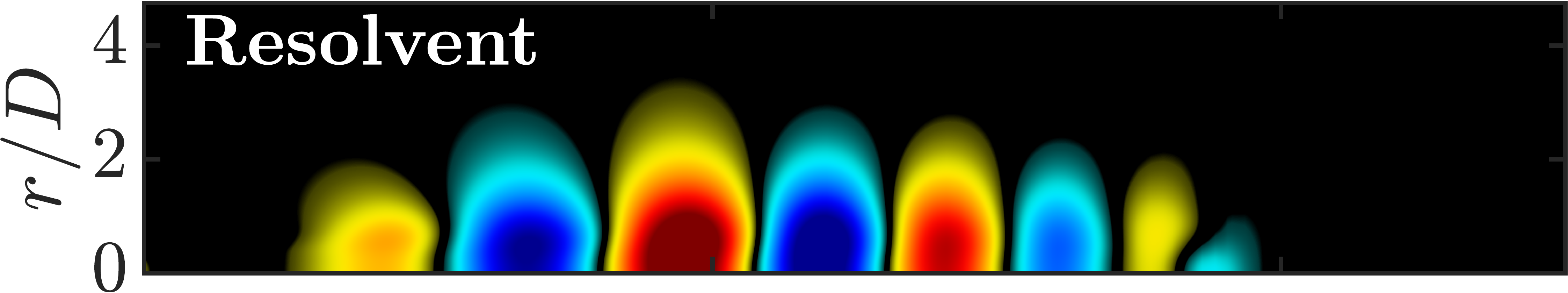} \\ 
\includegraphics[width=0.4\textwidth]{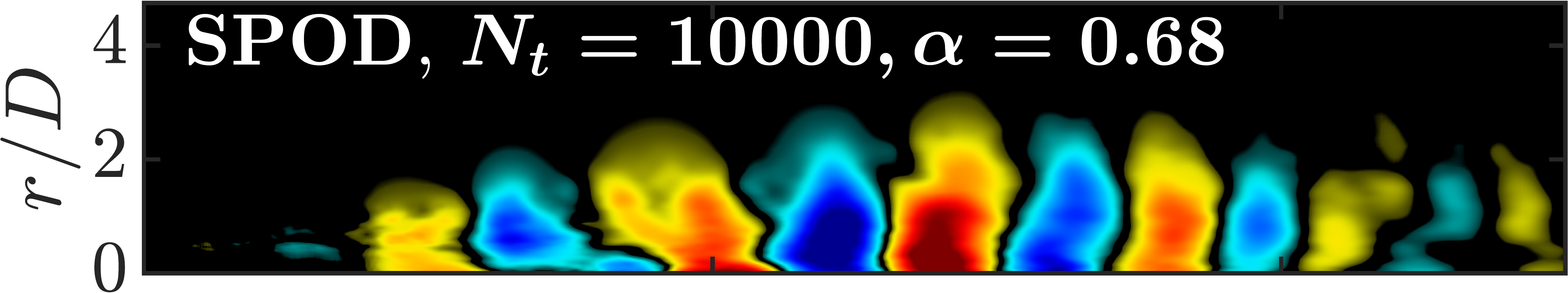}   \\
\includegraphics[width=0.4\textwidth]{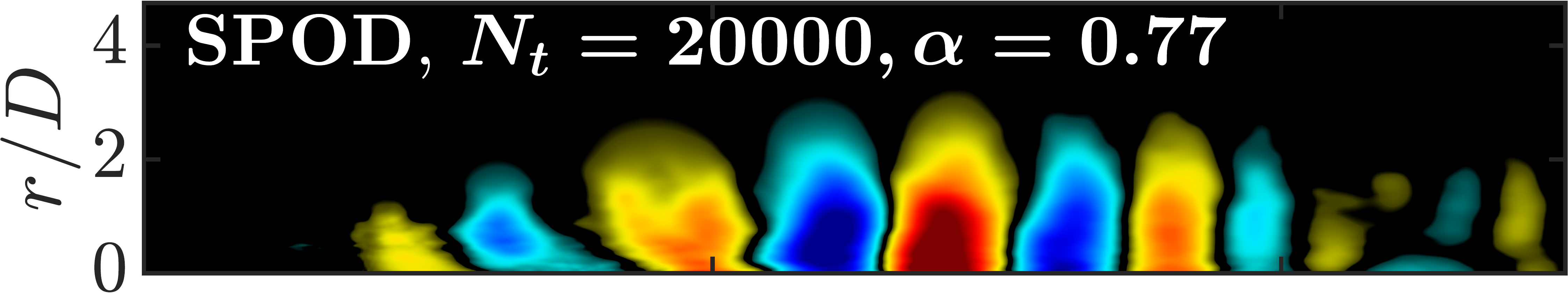} \\ \includegraphics[width=0.4\textwidth]{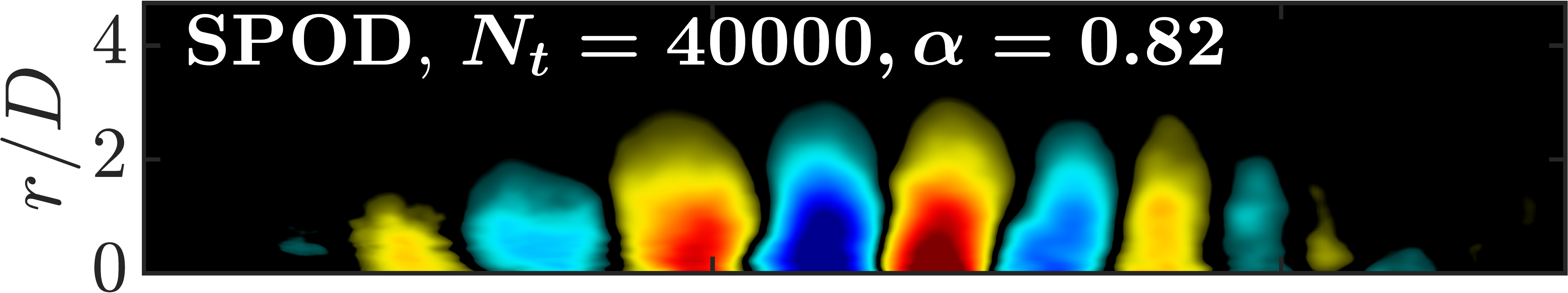}  \\
\includegraphics[width=0.4\textwidth]{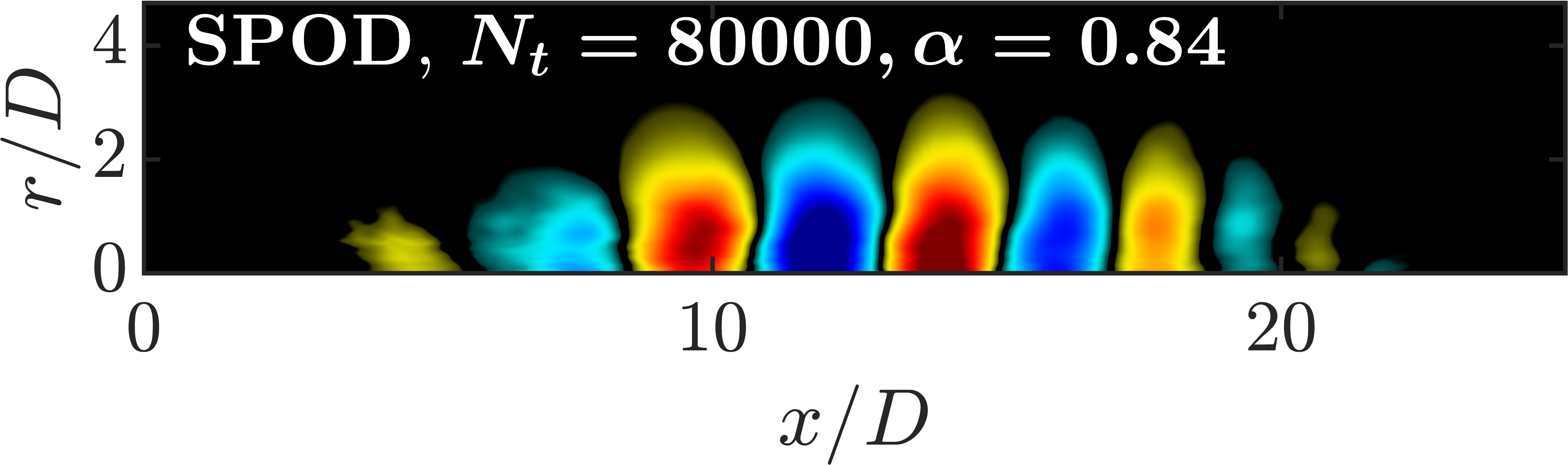}  \\
\caption{Dominant resolvent and SPOD modes at $St = 0.1$ for varying $N_t$ with $\Delta St = 0.025$. }
\label{fig:jetoptimalmodes}
\end{figure}

Lastly, in figure \ref{fig:jetoptimal} (c), we show the alignment as a function of the data length and $\Delta St = 0.025$. We also display the standard deviation of the alignment for $N_t = 10000$, where we find that the alignment is sensitive and varies greatly from one set of snapshots to another. We also find that the alignment increases substantially with increasing $N_t$ (particularly at lower frequencies). This shows that the alignment metric requires a significant amount of data to converge, even once the modes look visually similar, which we show for $St = 0.2$ in figure \ref{fig:jetoptimalmodes}.  Thus, due to the high variance and slow convergence of the alignment metric, caution is advised when inferring results from (or, in particular, comparing between) alignments. With respect to the modeling of turbulent flows, and in particular turbulent jets, the substantial increase in alignment with increasing $N_t$ (with the alignment increasing by up to $\approx 0.2$) demonstrates that resolvent analysis can model coherent structures even better than previously determined  \citep{pickering2021optimal}. \vsv

\vspace{-5mm}
\section{Conclusion}  \vsv
This work demonstrates that the choice of frequency resolution $\Delta f$ is vital to obtaining accurate SPOD modes, and the optimal choice critically depends on how rapidly the SPOD modes change in space at adjacent frequencies. We show that previously employed frequency resolution values are too high, which results in unnecessarily biased results at physically important frequencies. We demonstrate this using artificial data and data from a turbulent jet. We also develop a physics-informed adaptive frequency-resolution algorithm that provides substantially superior SPOD modes than the standard constant resolution method. Lastly, we demonstrate that the alignment metric widely employed to quantify the similarity between different SPOD/resolvent modes is highly sensitive to bias and the level of statistical convergence, and caution should be taken when inferring conclusions. Our results generalize to other spectral decomposition techniques such as Cyclostationary SPOD (CS-SPOD) \citep{heidt2023}, an extension of SPOD to flows with harmonic forcing such as externally forced flows or naturally harmonic phenomena such as vortex-shedding, Bispectral Mode Decomposition \citep{schmidt2020bispectral} which investigates triadic interactions, and Phase-Conditioned Localized SPOD \citep{franceschini2022identification} that investigates the evolution of high-frequency coherent structures on a low-frequency mean field. \\

\small{\textbf{Acknowledgments:} The authors gratefully acknowledge support from the United States Office of Naval Research under contract N00014-20-1-2311 with Dr. S. Martens as program manager and the Federal Aviation Administration under grant 13-C-AJFE-UI. This work was supported in part by high-performance computer time and resources from the DoD High Performance Computing Modernization Program. This work used Stampede2 at Texas Advanced Computing Center through allocation CTS120005 from the Advanced Cyberinfrastructure Coordination Ecosystem: Services \& Support (ACCESS) program, which is supported by National Science Foundation grants \#2138259, \#2138286, \#2138307, \#2137603, and \#2138296. }

\vspace{-3mm}
\bibliographystyle{elsarticle-num-names}
\bibliography{mybib.bib}
\end{document}